# Priority based Interface Selection for Overlaying Heterogeneous Networks

Mostafa Zaman Chowdhury and Yeong Min Jang
Department of Electronics Engineering
Kookmin University, Seoul 136-702, Korea
E-mail: {mzceee, yjang}@kookmin.ac.kr

*Abstract* — Offering of different attractive opportunities by different wireless technologies trends the convergence of heterogeneous networks for the future wireless communication system. To make a seamless handover among the heterogeneous networks, the optimization of the power consumption, and optimal selection of interface are the challenging issues for convergence networks. The access of multi interfaces simultaneously reduces the handover latency and data loss in heterogeneous handover. The mobile node (MN) maintains one interface connection while other interface is used for handover process. However, it causes much battery power consumption. In this paper we propose an efficient interface selection scheme including interface selection algorithms, interface selection procedures considering battery power consumption and user mobility with other existing parameters for overlaying networks. We also propose a priority based network selection scheme according to the service types. MN's battery power level, provision of QoS/QoE in the target network and our proposed priority parameters are considered as more important parameters for our interface selection algorithm. The performances of the proposed scheme are verified using numerical analysis.

## 1. Introduction

The tremendously increasing of the use of wireless networks for various applications has been seen in the recent years and it will continue in the future. Different wireless technologies have been developed due to these huge demands, varieties of user types, and varieties of user's requirement. The future wireless networks will be the convergence of these heterogeneous networks. These network technologies vary widely in terms of bandwidths, Quality of Service (QoS) provisioning, security mechanisms, price, coverage area and etc. Suppose, the complementary characteristics of WLANs and Universal Mobile Telecommunications System (UMTS) based cellular networks make them attractive for integration. This integration offers the best of both technologies. Thus, a mobile node (MN) with multiple wireless interfaces has become increasingly popular in recent years [1]. In heterogeneous overlay network, the MN can select one interface that is best or suitable in terms of price, QoS, Quality of Experience (QoE), throughput or other parameters as required. During connection, changes in the availability or characteristics of an access network may result in a situation where already established connections should be moved from one interface to another. This change of interface should be performed efficiently and seamlessly. The goal of the next generation network is to integrate multiple wireless access technologies to provide seamless mobility for the mobile users with high-speed wireless connectivity [2]. Seamless handover, resource management, and CAC to support QoS and multiple interface management to reduce power consumption in mobile terminal are the most important issues for the any multiple overlaying networks. Also, a MN, especially a battery-operated device with multiple wireless interfaces, power consumption is one of the critical problems [1].

Traditionally for horizontal handovers, only signal strength and available bandwidth are used as handover decision parameters. Also for traditional overlay network, the handover decision depends on several parameters like the signal strength, the available bandwidth, the price of the link, the security level, and the coverage radius [3]. As the user mobility and power of the battery are very important issues, these parameters should be considered as important parameter for suitable interface selection. Also for different applications, different parameters are important. Hence, for an efficient interface selection scheme, the weight of different parameters should be different for different applications. The power management issue can be added to IEEE 802.21 Media Independent Handover (MIH) [4], [5] for interface selection in overlay network. For our proposed interface selection algorithm, user mobility, MN's battery level and provision of QoS/QoE in the target interface have been considered as important parameters with other existing parameters for our interface selection scheme. The consideration of user mobility reduces some unnecessary handovers. Thus the proposed scheme will provide less interruption and better QoS as required by various applications in the wireless communication system.

The rest of the paper is organized as follows. Section 2 provides the related study about the power management issues in IEEE 802.21 MIH and interface selection schemes. Our proposed interface selection scheme is presented in Section 3.

In Section 4, we demonstrate the numerical results for the proposed algorithm. We give our conclusion in Section 5.

## 2. Related Works

Figure 1 shows one example that the users can be connected with the multiple interfaces. However, the battery power consumption for the multiple interfaces is more than that of use just single interface. Also, the access of different network interface causes different level of power consumption. Thus, an efficient mechanism and algorithm are needed for the selection of a best interface.

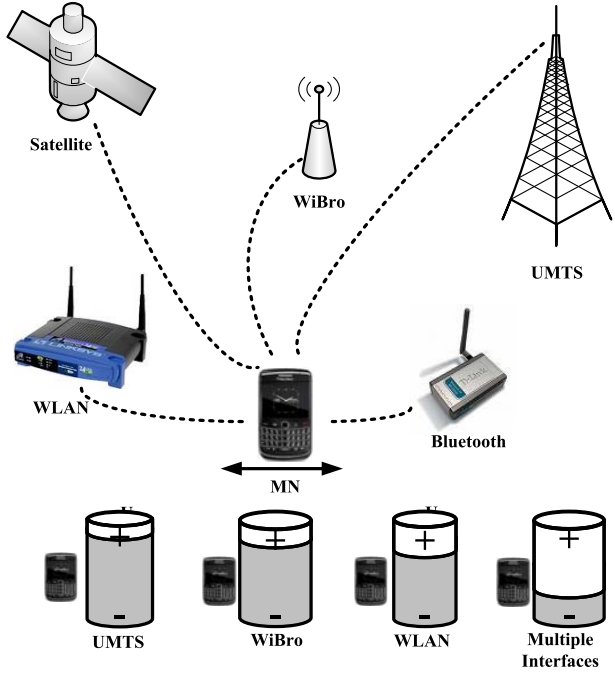

**Figure 1.** User connected with the multiple interfaces and effect of battery power consumption for the use of different interfaces

For the existing interface selection mechanism, several parameters are used for handover decision matrix. The interface selection procedure is a Multiple Attribute Decision Making (MADM) problem where alternative options are possible by multiple numbers of links (interfaces). Best handover decision depends on how the parameters are selected and how these parameters are used for interface selection algorithm. There are several works already done for this area. Different researchers [1]-[3], [5]-[9] assume different parameters for their interface selection algorithm but no one assume the status of MN's battery or battery profile, user mobility, and provision of QoS/QoE in the target interface as their interface selection algorithm. The authors in [3] used six parameters for each interface; signal strength, bit rate, power consumption, price, coverage and security. They made weight vector or profile for the interface selection algorithm. The Simple Additive Weighting (SAW) and Weighted Product proposed in [3] for the measurement of each property. They didn't consider user mobility and battery profile for their interface selection algorithm.

Authors in [10] proposed cost function based model for interface selection algorithm. Signal strength (*s*), cost of using the network access technology (*c*) and client power consumed for the particular access technology (*p*) are used as input parameters for their algorithm. They used just linear equation for the network selection algorithm. According to their algorithm, the score function (*SF*) for interface selection decision is

$$SF_i = (w_s * f_{s,i}) + (w_p * f_{p,i}) + (w_c * f_{c,i}) \quad (1)$$

According to [9] the power consumption for a specific application in WLAN $C_w$ and power consumption in UMTS $C_U$ are given by

$$C_w = (P_{tw}C_{tw} + P_{rw}C_{rw} + P_{lw}C_{lw} + P_{sw}C_{sw})T \quad (2)$$

$$C_U = (P_{tu}C_{tu} + P_{ru}C_{ru} + P_{pu}C_{pu} + P_{su}C_{su})T \quad (3)$$

In equation (2) and (3) $C_{tw}$, $C_{rw}$, $C_{lw}$, and $C_{sw}$ represents the power consumption in transmit, receive, idle and sleep state respectively for WLAN, while $P_{tw}$, $P_{rw}$, $P_{lw}$, and $P_{sw}$ are the probabilities of being in any of the respective communication state. $C_{tu}$, $C_{ru}$, $C_{su}$, and $C_{pu}$ represent the power consumption in transmit, receive, signaling and power-saving state respectively for UMTS, while $P_{tu}$, $P_{ru}$, $P_{su}$, and $P_{pu}$ are the probabilities of being in any of the respective communication state. Hence, power consumption also depends on different mode of operations.

## 3. Proposed Network Selection Scheme

### 3.1 Proposed Interface Selection Algorithm

The interface selection mechanism considers different parameters to select the best interface. The weight of all the parameters should not same to make a best selection. The algorithm should make a big weight for some higher priority parameters to emphasis those parameters. The traditional existing network selection algorithms [2]-[4], [7], [8] do not consider different priority parameters for different applications. They also do not consider the user mobility and normal or battery power saving mode for the interface selection algorithm. In our proposed algorithm we divide all the *m* number of interface selection parameters into two groups. One group takes more priority than another group for interface selection decision. The weight ($w_i$) of each parameter is different. We have *N* number of available interfaces. Suppose, battery power level ($P_{bat}$) and other *M* number of parameters for interface selection. Among the *M*, *q* number of parameters have less priority than other remaining (*M-q*) parameters, then the score (*S*) of the measurement is presented as

$$S = f_1(P_{bat}) * [f_2(w_1, w_2, ... w_q) + f_3(w_{q+1}, w_{q+2}, ... w_M)] \quad (4)$$

For the network selection algorithm, current level or status of MN's battery condition should be considered as well with other traditional parameters. We introduce some new parameters compared to traditional scheme. We consider three factors such as power saving/normal mode, low priority parameters, and higher priority parameters by three functions in (4). The function for the power saving/ normal mode enhances the score to select the lower power consuming interface. The function for higher priority parameters enhances the score to select the interface which provide better service in terms of higher priority parameters. For our proposed algorithm we consider battery power level and other eight parameters *(M=8)*; signal strength, throughput, power consumption, cost, cell coverage, QoS/QoE level, security, and user's mobility. Equation (4) measures the total score for each interface. We consider two modes of operation for network selection. Only for the power saving mode operation, we consider the MN's battery condition. For normal mode, we don't consider the MN's battery condition.

Suppose $w_m$ and $S_m$ indicates the weight and scaling factor of $m^{th}$ interface selection parameter respectively. $P_{bat}$ indicates the battery power level of the MN, then the score of $i^{th}$ interface among *N* interfaces can be calculated as following procedures:

The weight of each parameter is

$$0 \leq w_m \leq 1 \quad (5)$$

The summation of the scaling factors of all the interface selection parameters can be presented as

$$\sum_{m=1}^{M} S_m = 1 \quad (6)$$

From (5) and (6), we find that

$$0 \leq \sum_{m=1}^{M} w_m * S_m \leq 1 \quad (7)$$

The available *N* numbers of interfaces are ranked (R) according to the power consumption by each available interface. R is nothing but just ranking value of an interface in terms of power consumption. This ranking order may change if the position of the user is changed. The lower received signal by MN from the base station or access point causes higher battery power consumption of the MN. For lowest power consuming interface, *R=1* and highest power consuming interface *R=N*. By measuring the received signal (RSSI), the rank is done. The function $f_1(P_{bat})$ that is related to MN's battery is expressed by

$$f_1(P_{bat}) = \begin{cases} \dfrac{1}{\exp(1)} & normal\ mode \\ \dfrac{1}{\exp(1)log(10R)} & power\ saving\ mode \end{cases} \quad (8)$$

where *1≤R≤N* for *N* numbers of available interfaces.

The score for the lower priority interface selection parameters is calculated as

$$f_2(w_1, w_2, ..w_q) = \sum_{m=1}^{q} w_m * S_m \quad (9)$$

The linear function does not enhance the weight of higher priority parameters sufficiently. We use exponential function instead of linear function for the higher priority parameters. The exponential function is used to give more emphasis for the higher priority parameters. The weight for the higher priority interface selection parameters is calculated as

$$f_3(w_{q+1}, w_{q+2}, ..w_M) = \exp(\sum_{m=q+1}^{M} w_m * S_m) \quad (10)$$

In (8), we use exponential function to limit the total maximum score to one. Equations (4) to (10) are used to calculate the score of each candidate interface for the network selection. Equation (8) introduces battery power level condition in the interface selection algorithm. Thus, the MN will be operated in power saving mode if the battery level of the MN is insufficient or below a threshold level. Lower than threshold level means, the battery power level is going to be worst condition and thus the MN should select an interface that consumes lower power. Equation (8) is the function of power saving/normal mode. The term $\frac{1}{log(10R)}$ for the power saving mode reduces the total score by the factor $log(10R)$. Thus for the higher consuming interface, the score become less.

The impact of *R* in the interface selection algorithm may be changed by the designer as required. The lower priority and higher priority parameters are divided according the service type. In our proposal, we considered many parameters. All parameters are not equally important for every service. The lower priority and higher priority parameters are divided according the service type. In equation (10), the exponential function gives more emphasize on more important parameters to select that interface which provide better services in terms of those parameters. Based on the service nature and specified QoS requirement for various services the proposed classified priority parameters for interface selection are shown in Table 1. In case of more than two services, the algorithm will calculate the score for the individual service and then it will make an average to select the interface.

**Table 1.** Proposed classified priority parameters for interface selection

| Type of service | Lower priority parameters (m=1 to q) | Higher priority parameters (m=q+1 to M) |
|---|---|---|
| **Real-time voice** | Throughput, and power consumption. | Signal strength, cost, cell coverage, QoS/QoE level, security, and user's mobility. |
| **Streaming video** | Power consumption, cell coverage, QoS/QoE level, security, and user's mobility. | Signal strength, throughput, and cost. |
| **Command/ control** | Throughput, cost, power consumption, cell coverage, and user's mobility. | Signal strength, QoS/QoE level, and security. |
| **Background** | Signal strength, power consumption, cell coverage, QoS/QoE level, security, and user's mobility. | Throughput and cost. |

## 3.2 Interface Selection Procedure

The proposed interface selection steps are shown in Figure 2. The cross layer information for different network interfaces are collected, and then, these information and some pre-defined policies for interface selection are checked using proposed algorithm to make a best interface selection decision. The algorithm calculates total score for each interface. According to the result of the algorithm, all the available N interfaces are ranked. For example best interface is ranked as 1 and the worst one is ranked as N. Thus, MN will try to handover to best selected network. If resources are available in the best selected interface, then the MN handover to that interface otherwise it will try for the next ranked interface. This process will continue until (N-1) ranked interface.

The decision module collects information from user interface, battery profile, policy module, MIHF, and link information module. User interface provides information about the type of application, access technology, user's QoS/QoE requirement, and etc. Battery profile provides the information about battery power level. Policy module provides the pre-defined policies and interface selection algorithm. The link information module observes different layers condition and combine these information using cross layer optimization and then forward these information to decision module. The decision module selects a best interface according to the policies, and then forwards the decision to handoff module. The decision module also makes a rank for the available interfaces according to the total weight. The handover execution module executes the handover.

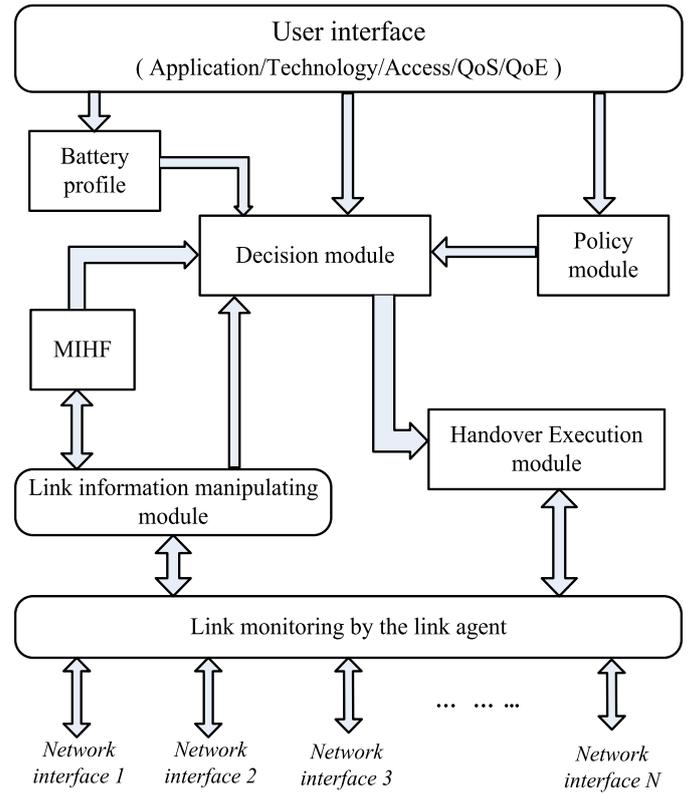

**Figure 3.** Proposed functional architecture and procedure for the interface selection

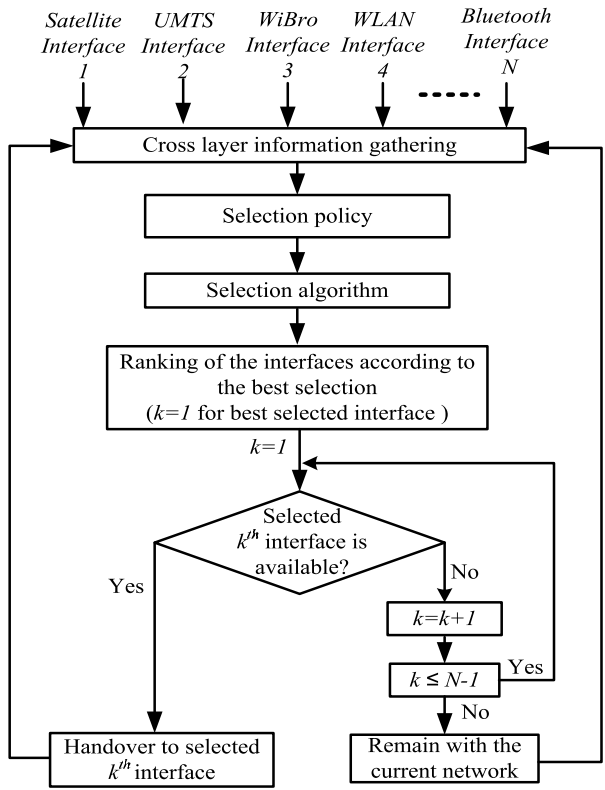

**Figure 2.** Proposed steps for the execution of interface selection scheme

Figure 3 shows the proposed functional architecture for the interface selection procedure. The link agent communicates with the physical layer. It provide the information about the detected all network interfaces.

## 4. Numerical Analysis

The performance of the proposed scheme is verified using numerical analysis in this section. We made several assumptions for the performance analysis. We consider background data traffic and real-time voice traffic in UMTS/WLAN overlaying networks. The Okumura-Hata model [11] for path loss is considered for the calculation in our numerical analysis. Table 2 shows all other basic assumptions for our analysis. Based on the importance of each parameter, the assumed scaling factor ($S_m$) of each parameter is given here. The weight ratio ($W_m$) of each interface is assumed according to their capabilities to provide level of services. The $S_m$ and $w_m$ may vary for different network condition and user requirement. In our assumption, we give emphasis on the cost and throughput. However, for power saving mode, we also give emphasis on battery power consumption. The best interface will be selected only when the resource in that interface is available. In our numerical analysis, we assume requested bandwidth is available in both the interfaces whenever the user requests for a call. From the path loss model, we calculate that the battery power consumption of a MN for UMTS interface is less than the WLAN interface if the distance of MN from UMTS BS is less than 1700m. However, the coverage area of UMTS base station (BS) is 2280m. If the distance of MN from WLAN AP is constant whatever the distance of the MN from the macro-cellular BS, the received signal strength from the WLAN AP can be considered as same value for every WLAN environment. Hence, the score of WLAN interface is almost constant for normal operating condition. But with the increasing the distance between UMTS BS and MN, the received signal level decreases and also battery power consumption increases. So, the total score of UMTS interface decreases with the increase of distance. The numerical results show that the total score of each interface for interface selection vary according to type of applications and environment. The results in four figures are taken for four different environments.

**Table 2.** Basic assumptions for the performance analysis

| Parameters for path loss model | | |
|---|---|---|
| Access network | Parameter | Assumption |
| UMTS | BS transmit power | 1.5 kw |
| | Path loss model (Okumura-Hata model for macrocell) | $L_p = 69.55 + 26.16 \log f_c - 13.82 \log h_b - a(h_m) + [44.9 - 6.55 \log h_b] \log d$ dB |
| | Height of BS | 100m |
| | Frequency | 900 MHz |
| | Receiver sensitivity | -130 dB |
| WLAN | AP transmit power | 100 mW |
| | Path loss model (Okumura-Hata model for microcell) | $L_p = 135.41 + 12.49 \log f_c - 4.99 \log h_b + [46.84 - 2.34 \log h_b] \log d$ dB |
| | Frequency | 2.5 G Hz |
| | Height of AP | 2m |
| | Coverage area | 30 m |
| Assumptions for weight parameters | | |
| Parameter | Scaling factor ($S_m$) | Weight ratio ($w_m$) |
| Cost | 0.35 | UMTS(0.1): WLAN(1) |
| Throughput | 0.15 | UMTS(0.1): WLAN(1) |
| QoS/QoE | 0.1 | UMTS(1): WLAN(0.25) |
| Mobility | 0.1 | UMTS(1): WLAN(0.01) |
| Signal strength | 0.1 | Depends on the distance between MN and UMTS BS |
| Power Consumption | 0.08 | Depends on the distance between MN and UMTS BS |
| Security level | 0.07 | UMTS(1): WLAN(0.25) |
| Cell coverage | 0.05 | UMTS(1): WLAN(0.01) |

Figure 4 shows the total score of WLAN and UMTS interfaces both for the background data traffic and real-time voice traffic in normal operating mode. As the distance between the WLAN AP and the MN is considered constant, the total score for WLAN interface is constant both for the data and voice services. It shows that WLAN is better due to lower cost, higher throughput and better signal quality in home environment. A data or voice user in WLAN coverage area where the received signal from WLAN AP is very good condition, will select a UMTS interface only if the resource of WLAN is not available. For the zero user velocity, the choice of WLAN interface will be cost effective for the user. Our algorithm also selects the WLAN. The selection of the WLAN interface for both the voice and data users enhances the throughput and reduces the cost.

Whenever the battery level of the MN is not sufficient, it will operate on the power saving mode. Figure 5 shows the total score of interface selection whenever battery power level is not sufficient. At this moment saving of power is more important. Thus, the MN can connect with the wireless link for longer time using power saving mode. In this condition, the interface selection algorithm gives more emphasis on that interface which consumes less power. Figure 5 also shows that the score of UMTS interface became very small after reaching 1700 m distance between the MN and the UMTS BS. Because of cell edge, UMTS causes more power consumption. At this distance, the total score of WLAN interface became high to emphasis the WLAN interface. In the power saving mode of operation, UMTS is an optimal choice whenever the distance between the MN and the BS is small. Even lower throughput and higher cost causes by the UMTS interface for the user, but the selection of this interface reduces the battery power consumption if the distance between the user and the UMTS BS is less than 1700 m. However, the data user selects WLAN for the distance between 600 m and 1700 m because throughput is more important for data user compare to saving of power. The selection of WLAN for the distance more than 1700 m causes higher throughput, lower cost and saving of power for both the voice and data users.

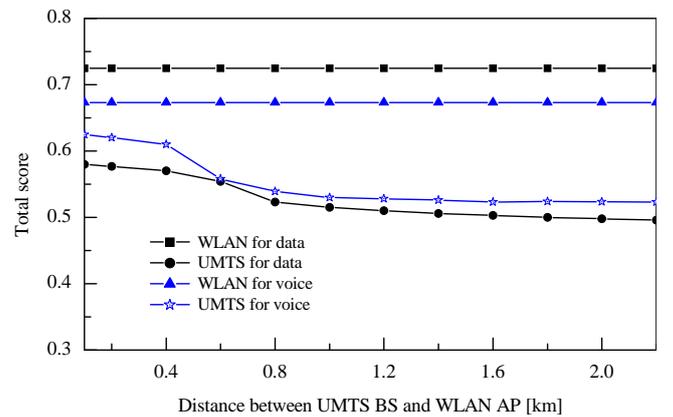

**Figure 4.** Total score of each interface whenever the MN has sufficient battery power level and changing UMTS signal environment. The distance between the MN and the WLAN AP is always 10 m. The user is assumed to be zero velocity.

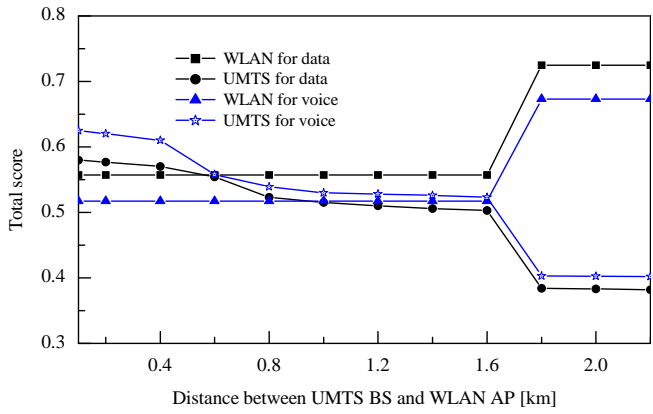

**Figure 5.** Total score of each interface whenever the MN has insufficient battery power level and changing UMTS signal environment. The distance between the MN and the WLAN AP is always 10 m. The user is assumed to be zero velocity.

Figure 6 shows the interface selection results for the changing WLAN signal environment with zero user velocity. In this condition, after 20 m distance between MN and WLAN AP, the voice user selects UMTS interface due to degraded WLAN signal level. However, data user still uses the WLAN interface due to lower cost and higher throughput. The data user can tolerate the QoS level. Thus, the data user selects the WLAN interface until the receiver is capable to receive the WLAN signal. Hence, the interface selection provides sufficient QoS level for the voice users. Here the selection of WLAN by data user causes higher throughput, lower cost. The selection of WLAN by voice user for less than 20 m distance between the WLAN AP and MN causes better signal quality, higher throughput, lower cost, and saving of battery power. However, the distance more than 20 m causes bad signal quality and causes higher power consumption. Hence the selection of UMTS interface can longer the battery lifetime of voice user for this condition.

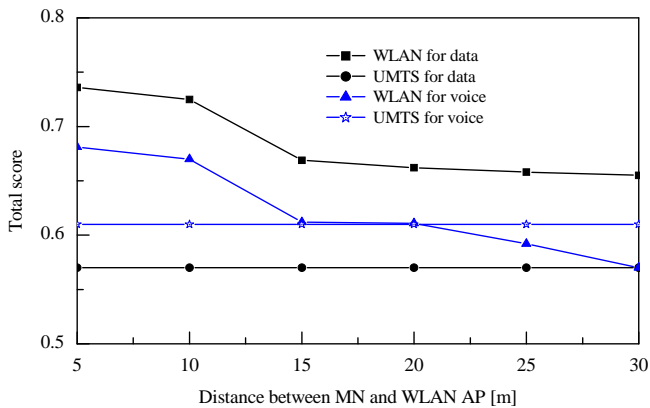

**Figure 6.** Total score of each interface whenever the MN has insufficient battery power level and changing WLAN signal environment. The distance between the MN and the UMTS BS is always 400 m. The user is assumed to be zero velocity.

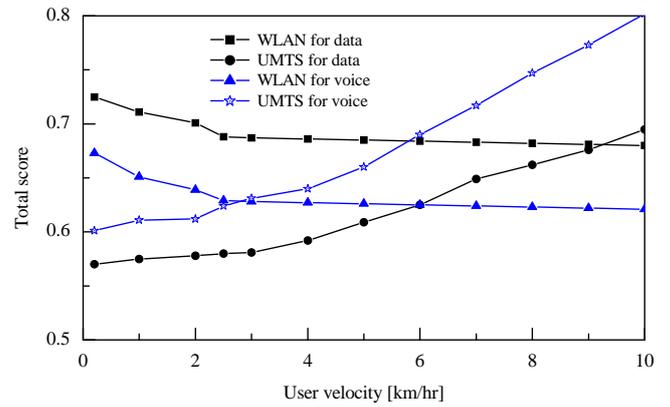

**Figure 7.** Total score of each interface whenever the MN has insufficient battery power level and changing the user mobility environment. The distance between the MN and the WLAN AP is always 10 m. The WLAN AP and the UMTS BS are 400m away.

The performances for the mobile users with different velocity are shown in Figure 7. The interface selection prominences the UMTS interface for the user with higher velocity. Especially for voice user, the score of the UMTS interface increases rapidly with increasing user velocity. The higher velocity causes more handovers in the system, higher dropping probability of call, and degrades the QoS level. The selection of the UMTS selection for higher velocity users improve the QoS/QoE level even though the selection of UMTS causes lower the throughput and higher cost.

All the numerical results in Fig. 4 to 7 show that the proposed scheme always selects the best interface to provide sufficient QoS with cheaper price whatever the environment. The proposed scheme chooses WLAN interface for normal conditions to increase the throughput. However, the proposed algorithm chooses the UMTS interface for the higher mobility condition. In Table 2, we just assume a user whose preferences of each parameters are represented by $S_m$. In normal condition, users always give more preference for cost and then throughput. In our assumption, we also gave more preference for cost and throughput by giving more scaling factor for them. Then, in our numerical results in Fig. 4 to 7, we proved that our algorithm is also cost effective and provide better throughput. For the battery power saving mode, our algorithm also emphasis that interface which consume less power. For the higher mobility case, our algorithm emphasis that interface which support higher mobility. These results are also feasible with the assumption. Thus our algorithm considers different condition. It can also be proved, suppose if we give more preference for security (for example in internet banking), then our algorithm will make more score for UMTS interface. Because, the UMTS provides better security level than WLAN. Thus, the proposed scheme is a promising scheme for the interface selection in multi-radio environment.

## 5. Conclusion

Multiple choices of interfaces are good opportunities to access multiple access networks with the suitable price and better QoS/QoE level as required. The main problems with the multi-mode operated MN are very high battery consumption and difficulties in the selection of best interface. In this paper we proposed priority based interface selection parameters. Based on the application types and user mobility, the more priority parameters are focused for the interface selection algorithm. The current battery power level has also been considered as the interface selection parameter for the interface selection algorithm. Thus, for lower battery level environment, the MN will operate in the power saving mode to select the low power consuming interface. We also considered QoS and QoE level that can be provided by target network in the interface selection algorithm. The proposed functional architecture for interface selection can provide a best handover decision for overlaying network. The numerical results show that the proposed algorithm is capable to select appropriate interface in both the normal operating mode and power saving mode. The algorithm gives more preference for that that interface which can provide better services in terms the higher scaling factor parameter. The MN will support the seamless services for longer time by the proposed power saving mode operation. The reduction of unnecessary handovers for the higher velocity users will enhance the QoS level.

## Acknowledgement

This research was supported by the MKE (Ministry of Knowledge and Economy), Korea, under the ITRC (Information Technology Research Center) support program supervised by the IITA (Institute of Information Technology Assessment) (IITA-2009-C1090-0902-0019).

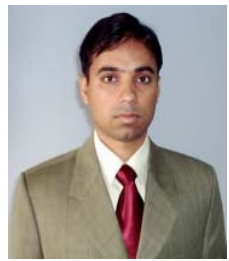
**Mostafa Zaman Chowdhury** received B.Sc. in EEE from Khulna University of Engineering and Technology (KUET), Bangladesh in 2002. Then, he joined as Lecturer and Assistant Professor in EEE department of KUET, in 2003 and 2006 respectively. He completed his MS in Electronics Engineering from Kookmin University, Korea in 2008. Currently he is continuing his Ph.D. studies in Kookmin University. His current research interests focus on convergence networks, QoS provisioning, mobility management, and femtocell networks.

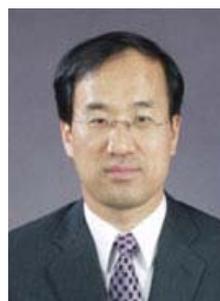
**Yeong Min Jang** received the B.E. and M.E. degree in Electronics Engineering from Kyungpook National University, Korea, in 1985 and 1987, respectively. He received the PhD degree in Computer Science from the University of Massachusetts, USA, in 1999. He worked for ETRI between 1987 and 2000. Since Sept. 2002, he is with the School of Electrical Engineering, Kookmin University, Korea. His research interests are IMT-advanced, radio resource management, and convergence networks.